\documentclass[twocolumn,showpacs,amsmath,amstex,amssymb,mathfonts,prl]{revtex4}
\usepackage{graphicx,bm,units}

\begin{document}

\title{Topological phonon modes and their role in dynamic instability of microtubules}

\author{Emil Prodan$^1$ and Camelia Prodan$^2$}
\address{$^1$Department of Physics, Yeshiva University, New York, NY 10016  \\
$^2$Department of Physics, New Jersey Institute of Technology, Newark, NJ 07102}

\begin{abstract}
Microtubules (MTs) are self-assembled hollow protein tubes playing important functions in live cells. Their building block is a protein called tubulin, which self-assemble in a particulate 2 dimensional lattice. We study the vibrational modes of this lattice and find Dirac points in the phonon spectrum. We discuss a splitting of the Dirac points that leads to phonon bands with non-zero Chern numbers, signaling the existence of topological vibrational modes localized at MTs edges, which we indeed observe after explicit calculations. Since these modes are robust against the large changes occurring at the edges during the dynamic cycle of the MTs, we can build a simple mechanical model to illustrate how they would participate in this phenomenon.
\end{abstract}


\date{\today}

\maketitle

MTs exist in every eukaryotic cell, being part of the  cellular cytoskeleton and  playing important roles in cell division  and intracellular transport. MTs are made of $\alpha$-$\beta$ tubulin protein dimers forming long protofilaments, which self-assemble into hollow tubes \cite{Desai:1997nx}. Under constant chemical conditions, the tubes switch randomly between growing and shrinking modes \cite{Fygenson:1994it}. This dynamic instability (DI) is essential to the functioning of the MTs.

In Ref.~\cite{Mahadevan:2005fe}, the MT's total energy was considered as function of the transversal and longitudinal curvatures of the tubulin sheet near the edge. It was proposed that the energy landscape has two minima separated by an energy barier, one favoring a cylindrical and the other a trumpet shape. Ref.~\cite{Mahadevan:2005fe} proposed that, during growth, the MTs reside near the first minimum, but somehow they start climbing until they overcome the energy barrier when they start sliding towards the second minimum, forcing the ends to open and the tubes to depolymerize.

\begin{figure}
  \includegraphics[width=7cm]{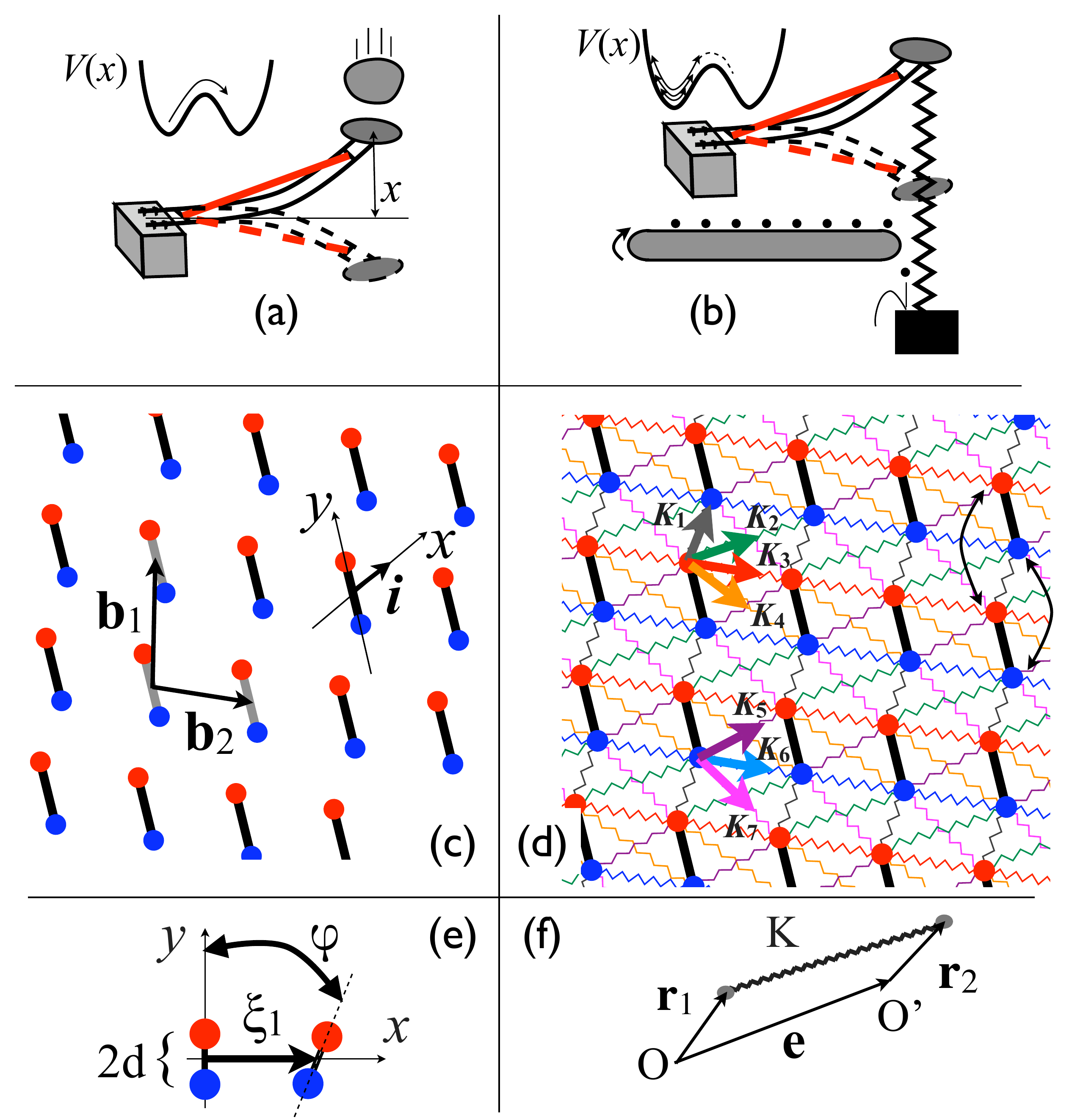}\\
  \caption{The system can be switched between two equilibrium configurations by: (a) delivering a large amount of energy at once or (b) by attaching a spring+weight and  releasing the beads one by one.(c) The 2D lattice of dimers of primitive vectors ${\bf b}_1$ and ${\bf b}_2$. At equilibrium, the dimers are all parallel and orientated along an arbitrary direction. The $xy$ coordinate system indicates the plane of motion. The $y$ axis is along the dimer but the $x$ axis is arbitrary. (d) The interaction between the tubulin dimers is modeled by a network of springs. There are 7 distinct springs and corresponding $\hat{{\bf e}}$ unit vectors. We also indicate second order neighbors whose harmonic interaction is also considered. (e) The degrees of freedom for the $xy$ planar motion.  (f) Stretched configuration of a spring with the ends displaced by ${\bf r}_{1,2}$.}
 \label{NewSetup}
\end{figure}

Fig.~\ref{NewSetup}(a) shows a device with a similar energy landscape. It is a bow made of two elastic roads hold together by two connectors and a rubber band stretched between them, so that it can slide freely between the two roads. The energy of the bow, as function of $x$ [Fig.~\ref{NewSetup}(a)], has two minima. We can switch between the minima by pushing hard on the free end, or by using a  succession of low energy actions. Indeed, let us attach a spring and a weight at the free end and aim small beads at the weight [Fig.~\ref{NewSetup}(b)]. The energy of the falling beads is stored in the harmonic motion of the weight, whose oscillation amplitude grows, forcing the bow at some point to climb the energy barrier and switch its configuration \cite{media}.

The spring+weight can represent a vibrational mode localized at the MT's edge and the beads a sequence of synchronized GTP hydrolyses. In ordinary lattices, the edge modes can easily disappear if changes occur at the edge. The edge mode mentioned above must be resilient, of special nature, because the properties of the edge vary wildly during DI. Switching for a moment from phonons to electrons, we point out that an entirely new class of materials, called topological insulators, has been discovered recently \cite{Kane:2005vn,Bernevig:2006ys,Konig:2007zr}, with the remarkable property that electronic states appear near any edge that is cut into such material. This is a consequence of the unusual {\it bulk} properties of the materials. For this reason, the electronic edge states cannot be destroyed by any chemical, mechanical, etc.,  treatment of the edge \cite{Prodan:2009lo,Prodan:2009mi}.

Using a realistic model, we demonstrate that the bulk phonon spectrum of the tubulin sheet can display the same unusual properties seen in the topological insulators. Based on a well established connection between the  bulk and the edge properties \cite{Prodan:2009lo}, we conclude that topological phonon states can appear at the MT's edges, which we indeed observe after explicit calculations. We advance the hypothesis that these topological edge modes play a role in the MT's DI, similar to that of the spring+weight in the mechanical model of Fig.~\ref{NewSetup}(a-b).

We model the tubulin sheet as a 2D lattice of rigid dimers [Fig.~\ref{NewSetup}(c)] with harmonic interactions [Fig.~\ref{NewSetup}(d)]. The system can support a variety of complex motions, but here we want to single out a particular motion that displays the interesting features mentioned above. Therefore, we restrict this study to propagating oscillatory motions that involve displacements and rotations of the dimers in the arbitrarily chosen, but fixed $xy$ local planes [Fig.~\ref{NewSetup}(c)], with the two degrees of freedom explained in Fig.~\ref{NewSetup}(e). We assume that the degrees of freedom left out are involved in oscillatory motions that either occur in different ranges of frequencies or couple weakly with the oscillatory motions that we just singled out.  

The elastic energy of a spring whose ends are displaced as in Fig.~\ref{NewSetup}(f) is $\frac{1}{2}K[\hat{{\bf e}} ({\bf r}_1$-${\bf r}_2)]^2$ [$\hat{{\bf e}}$=${\bf e}/|{\bf e}|$], if we retain only the quadratic terms. Therefore, the elastic energy for small oscillations stored in the network of springs is:
\begin{equation}
\begin{array}{c}
V=\nicefrac{1}{2} \sum_{\bf R} \left \{K_1[\hat{{\bf e}}_1 ({\bf r}'_{{\bf R}+{\bf b}_1}-{\bf r}_{\bf R})]^2 \right .  \\
+ K_2[\hat{{\bf e}}_2 ({\bf r}'_{{\bf R}+{\bf b}_1+{\bf b}_2}-{\bf r}_{\bf R})]^2  
+K_3[\hat{{\bf e}}_3 ({\bf r}_{{\bf R}+{\bf b}_2}-{\bf r}_{\bf R})]^2  \\
 +K_4[\hat{{\bf e}}_4 ({\bf r}'_{{\bf R}+{\bf b}_2}-{\bf r}_{\bf R})]^2
 +K_5[\hat{{\bf e}}_5 ({\bf r}_{{\bf R}+{\bf b}_2}-{\bf r}'_{\bf R})]^2  \\
 \left . +K_6[\hat{{\bf e}}_6 ({\bf r}'_{{\bf R}+{\bf b}_2}-{\bf r}'_{\bf R})]^2+K_7[\hat{{\bf e}}_7 ({\bf r}_{{\bf R}+{\bf b}_2-{\bf b}_1}-{\bf r}'_{\bf R})]^2 \right \}.
\end{array}
\nonumber
\end{equation}
The displacements ${\bf r}$ (${\bf r}'$) of the red (blue) beads can be written in terms of $\xi^{1}$ and $\varphi$  shown in Fig.~\ref{NewSetup}(c):
${\bf r}$=$(\xi^1$$+$$\xi^2) {\bf i}$ and ${\bf r}'$=$(\xi^1$$-$$\xi^2) {\bf i}$, where $\xi^2$=$d \varphi$. Moreover, since the dimers' moment of inertia is $I$=$2dM$, the kinetic energy is: $T=\sum_{\bf R} M \{(\dot{\xi}^1_{\bf R})^2+(\dot{\xi}^2_{\bf R})^2  \}$.

The equations of motion for the Lagrangean $L$=$T$$-$$V$, together with the ansatz: $\xi^{1,2}_{\bf R}$=$\frac{1}{2\pi}\mbox{Re}\int  e^{i{\bf k}\cdot {\bf n}-i\omega t}  \alpha^{1,2}_{\bf k} d{\bf k}$, where ${\bf n}$=$(n_1,n_2)$ denotes a dimer's place in the lattice, lead to the following equation for the normal modes:
\begin{equation}\label{NormalModes}
 \left (
\begin{array}{cc}
\epsilon_1({\bf k})-M\omega^2 & w({\bf k})  \\
\bar{w}({\bf k}) & \epsilon_2({\bf k})-M\omega^2
\end{array}
\right ) \left ( 
\begin{array}{c}
\alpha^1_{\bf k}  \\
\alpha^2_{\bf k}
\end{array}
\right ) =0,
\end{equation}
where ${\bf k}$ is in the Brillouin torus $[0,2\pi]$$\times$$[0,2\pi]$. $\epsilon_{1,2}({\bf k})$ are given in Table~\ref{tab} and [$\tilde{K}$=$K(\hat{{\bf e}}\cdot {\bf i})^2$]:
\begin{equation}
\begin{array}{c}
w({\bf k})=(\tilde{K}_3 -\tilde{K}_6 )[1-\cos k_2 ] 
+i(\tilde{K}_4  - \tilde{K}_5 ) \sin k_2  \\+i \tilde{K}_1 \sin k_1+  i \tilde{K}_2  \sin (k_1+k_2)
+i \tilde{K}_7  \sin (k_1-k_2).
\end{array}
\end{equation}
Eq.~\ref{NormalModes} has solutions only if the determinant of the matrix is null and this happens only if $\omega$ takes the values:
\begin{equation}\label{spectrum1}
\omega^2_\pm({\bf k}) = \nicefrac{1}{M}[\epsilon({\bf k}) \pm \sqrt{|\Delta \epsilon({\bf k})|^2 +|w({\bf k})|^2 }].
\end{equation}
Here, $\epsilon({\bf k})$ and $\Delta \epsilon({\bf k})$=$\frac{1}{2}[\epsilon_1({\bf k})$$\pm$$\epsilon_2({\bf k})]$, respectively.

\begin{figure}
  \includegraphics[width=6cm]{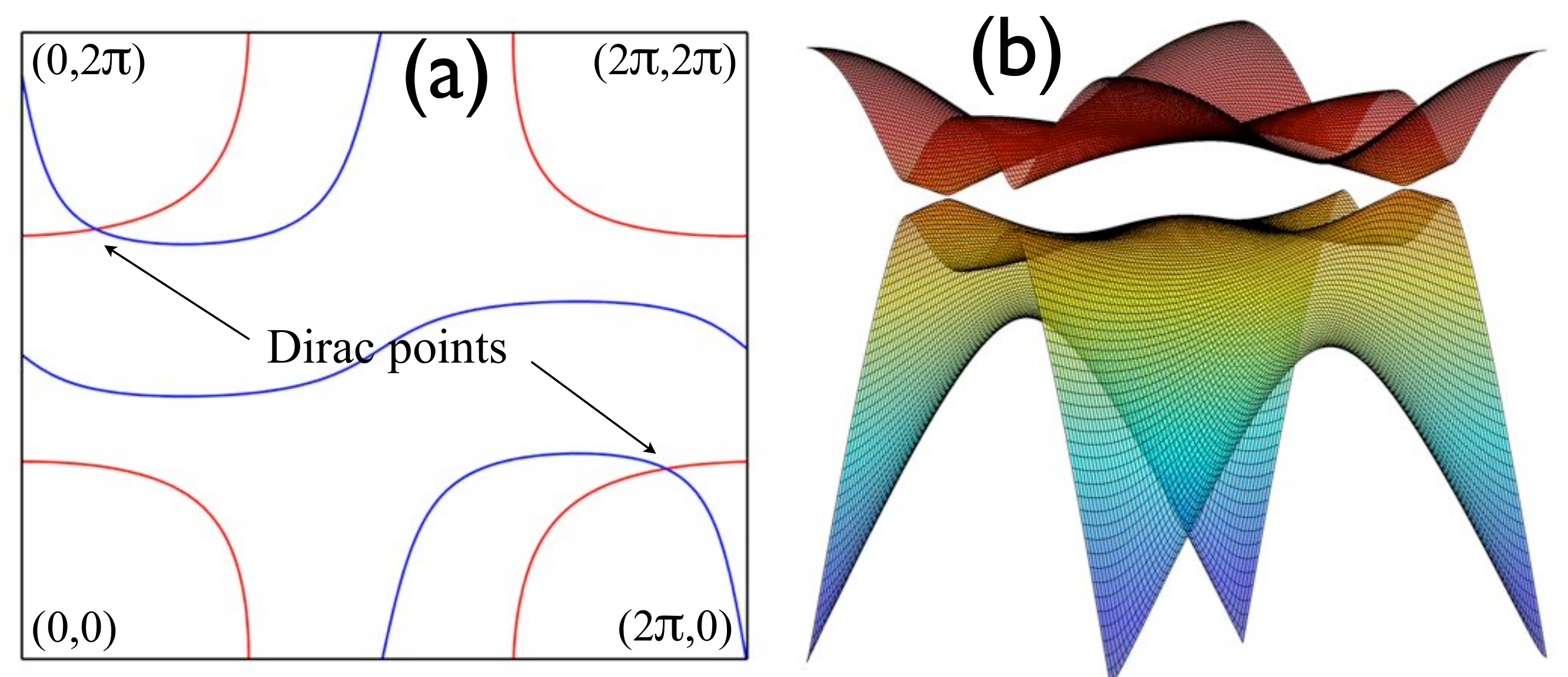}\\
  \caption{(a) The ${\bf k}$ values where $\Delta \epsilon({\bf k})$=0 (red) and $w({\bf k})$=0 (blue). (b) Plot of the phonon frequencies $\omega_\pm({\bf k})$. $\tilde{K}_{1-7}$ were fixed at: 1.1, 0.7, 0.27, 0.75, 0.325, 0.27, 0.75, and $M$=1.}
 \label{DiracPoints}
\end{figure}

Degeneracies $\omega_-({\bf k})$=$\omega_+({\bf k})$ occur whenever $\Delta \epsilon({\bf k})$ and $w({\bf k})$ are simultaneously zero. Due to the symmetries of the lattice, it is likely that $\tilde{K}_3$=$\tilde{K}_6$, in which case $w({\bf k})$ becomes purely imaginary. With this choice, we can view $\Delta \epsilon({\bf k})$=0 and $w({\bf k})$=0 as the equations of two closed loops on the Brillouin torus. There is a high probability for the two loops to intersect and, when it happens, degeneracies appears and the two phonon bands touch each other at isolated points (see Fig.~\ref{DiracPoints}). These special points are called Dirac points and their presence is a strong indication that our model has topological properties.

\begin{table}
\begin{tabular}{|c|c|c|c|c|}
	\hline
 &  $\cos k_1$    &   $\cos k_2$ &  $ \cos(k_1+k2)$  &  $\cos(k_1-k2)$   \\
	\hline
 $\epsilon_1({\bf k})-\epsilon_0$   & $- \tilde{K}_1 $  & $-\sum_{j=3}^{6}\tilde{K}_j$  & $- \tilde{K}_2 $ &  $-\tilde{K}_7 $ \\
\hline
 $\epsilon_2({\bf k})-\epsilon_0$    & $\tilde{K}_1 $ &  $\sum_{j=3}^6\delta_j \tilde{K}_j $  & $ \tilde{K}_2 $ & $ \tilde{K}_7 $ \\
	\hline
\end{tabular}\smallskip \\
\caption{$\epsilon_{1,2}({\bf k})$$-$$\epsilon_0$ are linear combinations of the first row, with coefficients specified in the next rows ($\delta_j$=1, except for $j$=3 and 6 when $j$=$-1$, and $\epsilon_0=\sum_{j=1}^7 \tilde{K}_j $.)}
\label{tab}
\end{table}

We now start the discussion of the bulk topological properties. If we allow a small difference between $\tilde{K}_3$ and $\tilde{K}_6$ or include additional terms in the Lagrangean, the Dirac points split and the bands separate. In this case, one can define the Berry curvature:
\begin{equation}
F_\pm({\bf k})= (2\pi i)^{-1} \text{Tr}\{ \hat{p}_\pm ({\bf k}) [\partial_{k_1} \hat{p}_\pm({\bf k}),  \partial_{k_2} \hat{p}_\pm({\bf k})] \}, 
 \label{curvature}
\end{equation}
where $\hat{p}^{mn}_\pm({\bf k})$=$\alpha_{\bf k}^m \bar{\alpha}_{\bf k}^n$ are the projectors onto the normal modes and $[,]$ denotes the usual commutator. The integrals $\int F_\pm({\bf k}) d^2 {\bf k}$ over the Brillouin torus are always integer numbers, called Chern numbers, which remain invariant under continuous deformations of the Lagrangean, unless the phonon bands touch.

When the Chern numbers are non-zero, the system acquires special topological properties \cite{Haldane:1988dp}. We show that this is possible in our model. As long as the Lagrangean is symmetric under the time reversal operation, the Chern numbers are zero. However, general quadratic Lagrangeans should also contain terms that involve products between velocity and displacement. Such terms break the symmetry mentioned above, thus allowing for non-zero Chern numbers. They can appear because of some weak magnetic properties of the tubulin proteins or because of the medium surrounding the lattice. A fairly general form of such terms will be:
\begin{equation}\label{vprime}
V'= \frac{1}{2}\sum_{\bf R}\sum_{m=1,2}\sum_{j=1,2}  \gamma_{m,j}\{ \dot{\xi}^m_{\bf R}  \ \xi^m_{{\bf R}+{\bf b}_j}-\xi^m_{\bf R}  \ \dot{\xi}^m_{{\bf R}+{\bf b}_j}  \}.
\end{equation}
Note that a plus combination inside the accolades leads to a total time derivative, which can be ignored. We expect the strength of $V'$ to be small relative to the terms already included in the Lagrangean. The second neighbors interactions  indicated in Fig.~\ref{NewSetup}(d),
\begin{equation}
\begin{array}{c}
V''=\nicefrac{1}{2} \sum_{\bf R}\sum_\# K_8^\# [\hat{{\bf e}}_8\cdot ({\bf r}^\#_{{\bf R}-{\bf b}_1}-{\bf r}^\#_{\bf R})]^2, 
\end{array}
\end{equation}
where $\#$ means primed or un-primed, could have similar strength and it should be considered at this point.

After including $V'$ and $V''$ into the Lagrangean, the equation of the normal modes takes the same form Eq.~\ref{NormalModes}, but we must add [$m$=1,2]
\begin{equation}\label{addition}
\begin{array}{c}
(\tilde{K}_8+\tilde{K}_8')[1-\cos k_1] 
+ \frac{\omega}{\omega_0} \sum_{j=1,2} \gamma_{m,j} \sin k_j
\end{array}
\end{equation}
to $\epsilon_{1,2}$, respectively, and $(\tilde{K}_8$$-$$\tilde{K}_8')[1$$-$$\cos k_1]$ to $w$. Here, $\omega_0$ is a characteristic frequency that will be specified shortly. We now search for those values of the parameters that give non-zero Chern numbers. We consider the parameters used to generate Fig.~\ref{DiracPoints}, except that now we allow a small mismatch $\tilde{K}_3$=0.27 and $\tilde{K}_6$=0.18. In addition, we take $\tilde{K}_8^1$=0.018 and $\tilde{K}_8^2$=0.1 and $\gamma_{1,1}$=-$\gamma_{2,1}$=$-\gamma_{1,2}$=$\gamma_{2,2}$=$\gamma$. In {\it contradistinction} to the electronic topological insulators, the Dirac points cannot be split only by the time reversal breaking term $V'$, because $w$ ($\epsilon({\bf k})$) remains purely imaginary (real). One should also be aware that the equation for the normal frequencies is also more complicated since now we have terms proportional to $\omega^2$ and $\omega$. Therefore, the observation of a parameter range where the Chern numbers are non-zero represents one of the important findings of our work. 

\begin{figure}
  \includegraphics[width=6.0cm]{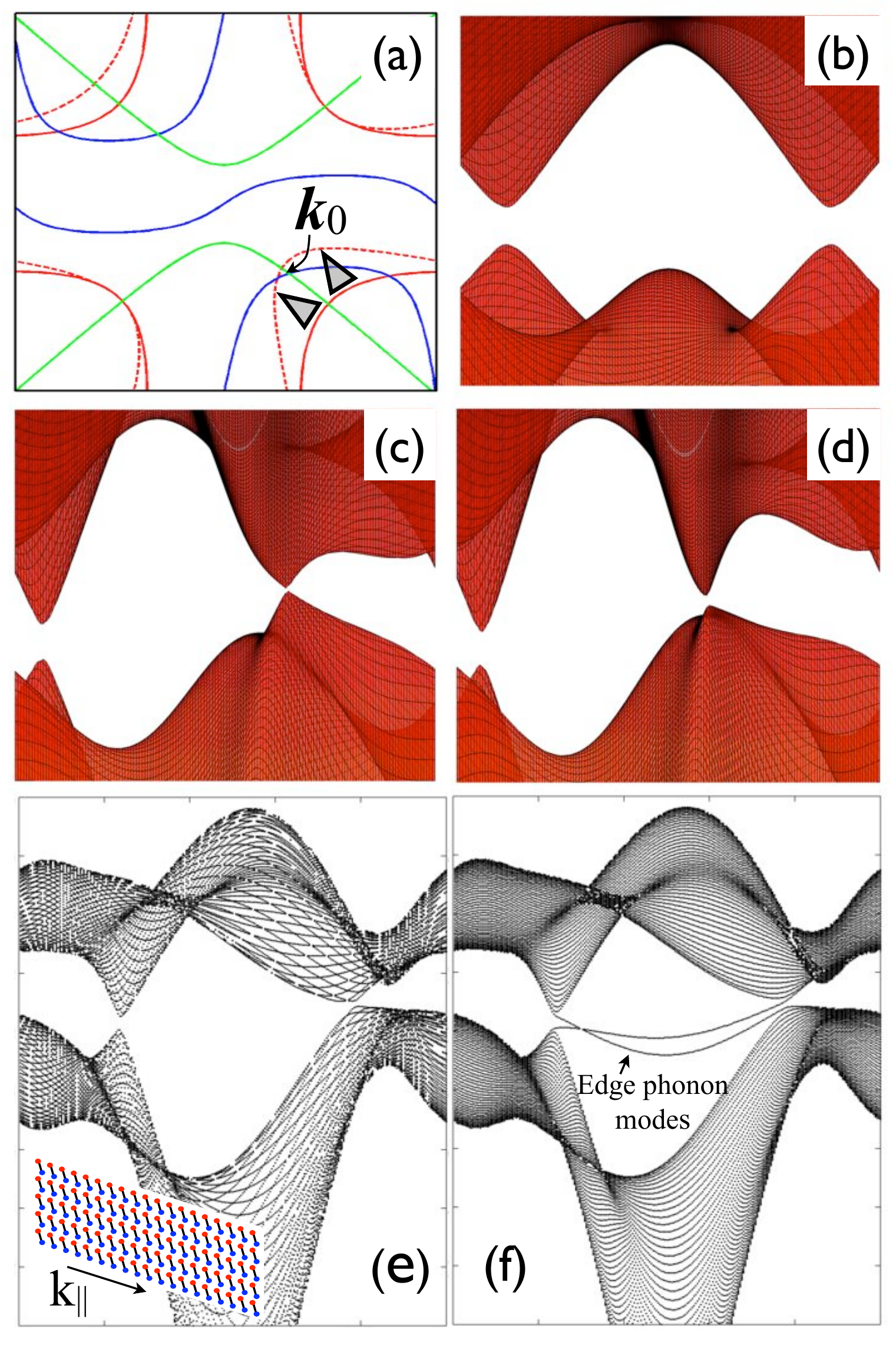}\\
  \caption{(a) The {\bf k} values where $\Delta \epsilon ({\bf k})$=0 (red), Re[$w({\bf k})$]=0 (blue) and Im[$w({\bf k})$]=0 (green). The full/dash lines are for $\gamma$=0/0.5. (b - d) The phonon spectrum for $\gamma$=0, 0.45 and 0.6. The last row shows the phonon spectrum of a ribbon as function of $k_{||}$ for (e) periodic and (f) open boundary conditions.}
 \label{chern}
\end{figure}

Let us set $\gamma$ to zero first. The Lagrangean is symmetric under time reversal, consequently, the Chern numbers are zero. We now increase $\gamma$, gradually introducing the term that breaks the time reversal symmetry. Even if we do so, the Chern numbers remain unchanged, unless the phonon bands touch and then separate again. We show that this is exactly what happens. Indeed, let us, for a moment, fix $\omega$ in Eq.~\ref{addition} to $\omega_0$, in which case the frequencies $\omega_\mp({\bf k})$ are given by same Eq.~\ref{spectrum1}. The bands touch if simultaneously: $\Delta \epsilon ({\bf k})$=0, Re[$w({\bf k})$]=0 and Im[$w({\bf k})$]=0. In Fig.~\ref{chern}(a) we plot the solutions for these equations, which are closed loops on the Brillouin torus. We color code them with red, green and blue, respectively. When $\gamma$ is increased from 0 to 0.5, the blue and green loops remain unchanged but the red loop moves in the directions indicated by the arrows, brushing over the intersection point ${\bf k}_0$ of the green and blue loops. Therefore there is a $\gamma$ (=$\gamma_0$) for which $\omega_-({\bf k}_0)$=$\omega_+({\bf k}_0)$. We now unfreeze $\omega$ in Eq.~\ref{addition} and take $\omega_0$=$\omega_\pm({\bf k}_0)$. This does not change the normal modes equation at ${\bf k}_0$, therefore the bands will still touch when $\gamma$=$\gamma_0$. In Figs.~\ref{chern}(b-c) we plot the actual phonon bands for three values of $\gamma$, showing how the bands touch and then separate as $\gamma$ is increased. We have also verified directly that the Chern numbers switch their values from 0 to $\pm 1$ when $\gamma$ is about $0.45$.

Whenever the Chern numbers are not zero, vibrational modes appear near any edge cut into the 2D lattice \cite{Prodan:2009lo}. We can now perform a routine phonon band calculation to observe these modes [$\gamma$=0.6 in these calculations]. We considered a tubulin ribbon along the direction of ${\bf b}_2$ and computed the phonon spectrum as function of $k_{||}$ using periodic and open boundary conditions (see Fig.~\ref{chern}(e-f)). Only in the last case one can see two distinct phonon bands separating from the bulk spectrum. The two bands are localized on the two edges of the ribbon. Also, each band connects the lower and upper parts of the bulk spectrum, therefore, no matter how one distorts the ribbon, the bands will always cross the direct band gap. 

At this point we established the existence of robust phonon edge modes and now we return to the mechanism of DI. According to the cap theory \cite{Desai:1997nx}, which we adopt here, the bulk of the microtubules is made of predominantly GDP-tubulin, while the edge displays a cap of a few rows of GTP-tubulin. The GTP and GDP bound tubulin prefer parallel and tilted orientations relative to the tube's axis, respectively. During polymerization, GTP-tubulins from solution attache to the edges and, at almost the same rate, GTP-tubulins from the back rows of the cap turn into GDP-tubulins. The newly formed GDP-tubulins would prefer a tilted orientation, but they are locked inside the lattice and therefore the tilt cannot occur and instead mechanical strain develops. One established view is that this strain energy is hold securely by the cap, but if the cap opens the strain energy is released and the MT depolymerize.

All DI models are sketchy about how the MT opens its cap. The cap could be forced open or simply shed off. In either case, the MT's edge must cross an energy barrier as described in Ref.~\cite{Mahadevan:2005fe}. The only energy available during DI are quanta of  12 kT released whenever GTP hydrolysis. The energy barrier discussed in Ref.~\cite{Mahadevan:2005fe} was never measured, but a rough yet fair estimate will place it around 50 kT. It is then evident that the MT must have a way of accumulating part of the 12 kT energy and concentrate it near the edge. Note that the strain energy is uniformly distributed within the bulk of the tube.

In our view, the sudden strain due to GTP hydrolysis gives a kick \cite{Howard:2003ve} to the edge mode, which can enhance or reduce the amplitude of the mode.  There is also dissipation from the edge mode into the bulk and surrounding medium, therefore, we cannot expect a slow, steady energy buildup at the edge. Instead, the MT needs a small number of lucky strikes, synchronized well enough to push the edge mode over the energy barrier. Assuming a 50 kT barrier and a 12 kT energy transfer to the edge mode during the lucky strikes, the MT needs about 5 consecutive lucky strikes to open the cap. The time till such strikes occur is random, which can explain the stochastic behavior of the catastrophe events. 

The existing calculations put the MT's vibrational modes in the frequency range from  1 MHz to few GHz and the ratio between their relaxation times and the period was estimate to be as high as $10^4$ \cite{Pokorny:2004gf}. The average time between consecutive tubulin dockings can be derived from the polymerization speed \cite{Fygenson:1994it} and is about $\tau_{\mbox{\tiny{av}}}$=0.02 s. The kicks, however, come from random GTP hydrolysis at the back of the cap and involve tubulins already attached to the edge \cite{Dimitrov:2008ul}. The average time between consecutive hydrolyses is comparable to $\tau_{\mbox{\tiny{av}}}$, but the processes themselves happen much faster and they are also much simpler than the docking of the tubulins. Therefore, the well synchronized kicks can occur within time intervals much smaller than $\tau_{\mbox{\tiny{av}}}$ and comparable to the phonon relaxation times.

Anticancer drugs can modify DI. Taxol, for example, increases the growth period without strengthening the cap or stiffening the bulk of the MT \cite {Needleman:2004ek}.  The effect is present even when there is one bound taxol per several hundred dimers. Equally puzzling, at these small concentrations, taxol stabilizes the MTs without changing the growth rate \cite{Jordan:2004cr}.  This means the energy flow from GTP hydrolysis during the growth period remains unchanged. It seems that the MTs edges are less efficient at harvesting energy in the presence of taxol. 

In our view, taxol modifies the bulk properties of the MTs, leading to a delocalization of the edge mode. The localization near the edge of the topological mode is strong when the bulk phonon bands are widely separated and becomes weaker as the bands come towards each other \cite{Prodan:2009lo}. In the mechanical model of Fig.~\ref{NewSetup}, delocalization is like attaching the weight away from the free end of the bow. But the further we attach them the harder it is to switch the bow's configuration and a larger number of lucky strikes is needed. Thus, in our picture, the duration of the MT's growth depends on the localization of the edge mode, which is controlled by the MT's {\it bulk} properties.

To conclude, we advanced the hypothesis that phonon edge modes play important role in the dynamical instability of microtubules. Using an explicit lattice model, we demonstrated the existence of topologically robust edge phonon modes and we advanced the hypothesis that the MT uses them to concentrate energy near the edge in order to open the cap. The bulk lattice model can be extended to include an anharmonic lattice model of the cap, with a double-well energy landscape, allowing us to simulate and observe the opening of the cap explicitly. This and designing experiments to test our hypotheses are underway.

%\bibliography{microtubules}

\end{document}